\documentclass[prl,aps,twocolumn,showpacs]{revtex4}
\usepackage{graphicx}
\usepackage{dcolumn}

\begin{document}

\title{Is Sr$_2$RuO$_4$ a triplet superconductor?}

\author{K. Machida}
\affiliation{Department of Physics, Okayama University,
Okayama 700-8530, Japan}
\author{M. Ichioka}
\affiliation{Department of Physics, Okayama University,
Okayama 700-8530, Japan}
\date{\today}

\begin{abstract}
The field dependence of the specific heat $\gamma(H)$
at lower temperatures in Sr$_2$RuO$_4$ is analyzed
by solving microscopic  Eilenberger equation numerically.
We find that systematic $\gamma(H)$ behaviors from
a concaved $\sqrt H$ to a convex $H^{\alpha} (\alpha>1)$ under
$H$ orientation change are understood by taking account  of the
Pauli paramagnetic effect. The magnetizations are shown to be consistent 
with it. This implies either a singlet pairing or
a triplet one with $d$-vector locked in the basal plane, which allows us to 
explain other mysteries of this compound in a consistent way.
\end{abstract}



\maketitle
Superconductors are classified into two distinctive groups,
either spin-singlet or spin-triple pairings.
While almost all superconductors, including high $T_{\rm c}$ cuprates 
belong to the former, the later is extremely rare and difficult to 
find. Only a few examples of superconductors are discussed 
for its possibility;
In a heavy Fermion material UPt$_3$ the identification of a 
triplet pairing has been firmly established~\cite{taillefer,machida}.
The observed multiple phase diagram in field ($H$) versus
temperature ($T$) plane, consisting of three phases A, B and C,
 is reasonably explained only in terms of triplet pairing.
This situation is similar to superfluid $^3$He where two subphases ABM and BW
are identified in pressure vs. $T$ plane~\cite{leggett}.
Knight shift (KS) experiment by NMR  has played a fundamental 
role to confirm the theoretical predictions in UPt$_3$. It was particularly crucial
that both field directions where KS is changed and unchanged 
below $T_c$ are found  experimentally~\cite{tou} as predicted~\cite{machida},
identifying the $d$-vector direction.

Sr$_2$RuO$_4$ is second prime candidate for 
a triplet pairing superconductor~\cite{maeno0}.
A variety of theoretical and experimental works have been devoted 
to establishing it, but it turns out after a decade of its discovery~\cite{maeno00}
that it is extremely difficult to identify the spin structure of a Cooper 
pair although the gap structure with line node is well established now. 
For example, it is pointed out that recent phase-sensitive experiments by Nelson {\it et al.}~\cite{nelson},
Kidwingira {\it et al.}~\cite{kid} and Xia {\it et al.}~\cite{xia}, all of which claim a triplet pairing,
are  also explained in terms of the singlet scenario by Zutic and Mazin~\cite{mazin}
and Mineev~\cite{mineev}.
The most direct and virtually only probe to detect its parity
is the KS experiment. In fact KS experiments using various nucleus,
such as $^{87}$Sr,$^{101}$Ru, $^{99}$Ru and $^{17}$O atoms, 
fail to pin down the spin 
direction of pairs, i.e. orientation of the $d$-vector
because of the invariance of KS for both field directions
of $c$- and $ab$-axes as low as $H=200G$~\cite{murakawa}. 
There is no field direction where KS
changes below $T_c$. Thus at present it is fair to say 
that the two scenarios either based on singlet and triplet pairings are 
still under debate. Note that the appearance of magnetic field below $T_c$
associated with  spontaneous  time reversal symmetry breaking 
observed by $\mu$SR experiment~\cite{luke} is explained equally by spin singlet 
scenario as well as triplet one~\cite{kirtley}.

We examine the parity issue in Sr$_2$RuO$_4$ through analyses
of the specific heat experiment by 
Deguchi {\it et al.}~\cite{deguchi0} under  various $T$ and $H$. There are several 
outstanding problems posed by this experiment,
whose understanding leads to a new clue for this debate.
One of the most interesting discoveries is why the field
dependence of the Sommerfeld coefficient $\gamma(H)=
{\rm lim}_{T\rightarrow 0}C/T$ ($C$ is the specific heat) in the basal plane shows
a concave curvature in spite of the existence of the line node gap.
Namely,  this is quite at odd because $\gamma(H)$ is  
expected to be a $\sqrt H$-like behavior with a convex curvature
due to line nodes, i.e. the so-called Volovik effect \cite{volovik}.
It is remarkable to see that the concave curve becomes
a Volovik $\sqrt H$ curve with a convex curvature when the 
direction of the applied field moves away only by a few degrees of angle $\theta$ from
the basal $ab$-plane (see inset (a) in Fig.3).
In addition to analyses of the specific heat data\cite{deguchi0} 
we also examine magnetization data\cite{tenya}
at low temperatures under a field.
We explain these experiments based on an idea that strong 
Pauli paramagnetic effect is important in the basal
$ab$ plane physics of Sr$_2$RuO$_4$ and establish
a consistent picture for its superconductivity.

We calculate the  vortex lattice state properties
by quasiclassical Eilenberger theory in the clean 
limit~\cite{ichiokaQCLd1}.
This framework is valid when $k_F\xi\gg 1$ ($k_F$ Fermi wave number and
$\xi$ coherent length),  which is satisfied by Sr$_2$RuO$_4$.
We include the paramagnetic effects 
due to the Zeeman term $\mu_{\rm B}B({\bf r})$.
The flux density of the internal field is $B({\bf r})$ and 
$\mu_{\rm B}$ is a renormalized Bohr 
magneton~\cite{adachi}.  
The quasiclassical Green's functions
$g( \omega_l +{\rm i} \tilde{\mu} B, {\bf k},{\bf r})$, 
$f( \omega_l +{\rm i} \tilde{\mu} B, {\bf k},{\bf r})$ and 
$f^\dagger( \omega_l +{\rm i} \tilde{\mu} B, {\bf k},{\bf r})$  
are calculated in the vortex lattice state  
by the Eilenberger equation 
\begin{eqnarray} &&
\left\{ \omega_n +{\rm i}\tilde{\mu}B 
+\tilde{\bf v}({\bf k}_{{\rm F}}) \cdot\left[
\nabla+{\rm i}{\bf A}({\bf r}) 
\right]\right\} f
=\Delta({\bf r})g, 
\nonumber 
\\ && 
\left\{ \omega_n +{\rm i}\tilde{\mu}B 
-\tilde{\bf v}({\bf k}_{{\rm F}}) \cdot\left[ 
\nabla-{\rm i}{\bf A}({\bf r}) 
\right]\right\} f^\dagger
=\Delta^\ast({\bf r})g  , \quad 
\nonumber 
\label{eq:Eil}
\end{eqnarray} 
where $g=(1-ff^\dagger)^{1/2}$, ${\rm Re} g > 0$, and the 
normalized Fermi velocity
$\tilde{\bf v}$ is introduced so that 
$\langle \tilde{\bf v}^2\rangle_{\bf k}=1$
where $\langle \cdots \rangle_{\bf k}$ indicates the Fermi surface average. 
The paramagnetic parameter is
$\tilde{\mu}=\mu_{\rm B} B_0/\pi k_{\rm B}T_{\rm c}$. 
We consider the $d$-wave pairing  
for a pairing function with line nodes on the two-dimensional (2D)
cylindrical Fermi surface.
The pair potential is selfconsistently calculated. 
The vector potential ${\bf A}$ for the internal magnetic field 
is selfconsistently determined by 
considering both the diamagnetic contribution of 
supercurrent  and the contribution of the paramagnetic moment.
We consider the large Ginzburg-Landau parameter $\tilde{\kappa}=20$.
The local density of states is given by
$N({\bf r},E)=N_{+1}({\bf r},E)+N_{-1}({\bf r},E)$ with 
$N_\sigma({\bf r},E)=\langle {\rm Re }
\{
g( \omega_l +{\rm i} \sigma\tilde{\mu} B, {\bf k},{\bf r})
|_{i\omega_l \rightarrow E + i \eta} \}\rangle_{\bf k}$
for each spin component $\sigma=\pm 1$.
We typically use $\eta=0.01$.
The density of states is obtained by its spatial average as 
$N(E)=\langle N({\bf r},E) \rangle_{\bf r}$, which is identified 
as the Sommerfeld coefficient $\gamma(H)$ in specific heat at lower $T$. 
Using the Doria-Gubernatis-Rainer scaling,
we calculate magnetization $M$ including diamagnetic
and paramagnetic contributions.
The details are found in Refs.~\cite{ichiokaQCLd1,adachi}.

The paramagnetic parameter $\tilde{\mu}\propto {H^{\rm orb}_{c2}/H_p}$,
which is a key parameter to analyze $\gamma(H)$,
is related to the ratio of the hypothetical orbitally limited upper critical field 
$H^{\rm orb}_{c2}$ and the Pauli limiting field $H_p={\Delta_0/{\sqrt 2}\mu_B}$
($\Delta_0$ is the gap amplitude at $T=0$).
$H_p$ is a material-specific bulk parameter independent of the field orientation
evidenced by nearly isotropic bulk susceptibility observed~\cite{maeno0}.
The angle-dependence of the paramagnetic parameter $\tilde{\mu}(\theta)$ comes 
through the factor: $H^{\rm orb}_{c2}(\theta)$. This orbital-limited
 $H^{\rm orb}_{c2}(\theta)$ is sensitive to the field orientation 
for highly anisotropic system such as in the present layered material; Sr$_2$RuO$_4$.

The reduction of $H_{c2}$
from $H^{\rm orb}_{c2}$
due to the paramagnetic effect is obtained by solving the Eilenberger equation as 
$H_{c2}(\tilde{\mu})={H^{\rm orb}_{c2}/
\sqrt{1+2.4{\tilde\mu}^2}}$.
This is derived originally in dirty limit $s$-wave case~\cite{saint}, but we confirm it
to be valid numerically in the present clean limit $d$-wave case too as seen from Fig. 1
where the calculated values are compared with this expression.

It is natural to consider that $H^{\rm orb}_{c2}(\theta)$ is 
described by the effective mass 
model, namely
$H^{\rm orb}_{c2}(\theta)/H^{\rm orb}_{c2\parallel ab}=1/\sqrt{\Gamma^2\sin^2\theta+\cos^2\theta}$
which simply embodies the fact that the orbital motion of electrons is 
determined by the directional cosine of the field to the basal plane.
The anisotropy $\Gamma={H^{\rm orb}_{{c2}{\parallel ab}}/ H^{\rm orb}_{{c2}{\parallel c}}}$
is an unknown parameter here. But it is assigned by the
requirement that the experimental $H_{c2}(\theta)$ be reproduced theoretically.
Namely, once $\Gamma$ is determined, the angle dependence of 
$H_{c2}(\theta)$  is automatically known through the angle dependence of the 
paramagnetic parameter $\tilde{\mu}(\theta)$, which controls the reduction of the 
upper critical field $H_{c2}$ from the ``hypothetical'' orbital-limited field $H^{\rm orb}_{c2}$.

\begin{figure}[!h]
\includegraphics[width=5cm]{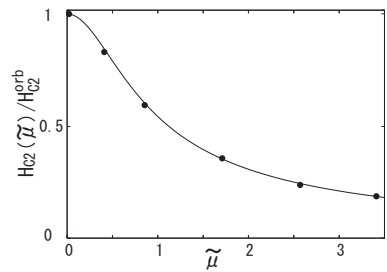}
  \caption{
Reduction of $H_{c2}(\tilde{\mu})$ as a function of $\tilde{\mu}$
evaluated by quasiclassical Eilenberger equation. The fitting curve is
described well by $H_{c2}(\tilde{\mu})/H^{\rm orb}_{c2}={1/
\sqrt{1+2.4{\tilde\mu}^2}}$.
}
 \label{fig:nchi}
\end{figure}

Having known the paramagnetic depairing effect on $H_{c2}(\tilde{\mu})$,
we can calculate the angle dependence of the observed $H_{c2}(\theta)$
where we take account of the fact that $\tilde{\mu}\propto {H^{\rm orb}_{c2}/ H_p}$
 is $\theta$-dependent through the factor
$H^{\rm orb}_{c2}(\theta)$ given above.
Thus we obtain 
${\tilde\mu}(\theta)={{\tilde\mu}_0/\sqrt{\Gamma^2\sin^2\theta+\cos^2\theta}}$
 with ${\tilde\mu}_0$ being the value at $\theta=0$.
 By combining these relations, we finally obtain the $\theta$ dependence of
 the observed $H_{c2}(\theta)$ as 
 $H_{c2}(\theta)=1/ \sqrt{\Gamma^2\sin^2\theta+\cos^2\theta+2.4{\tilde\mu}_0}$.
This takes account of both orbital- and paramagnetic depairing
effects simultaneously.
In order to reproduce the observed anisotropy $\Gamma^{\rm obs}=20$,
we find ${\tilde\mu}_0=3.41$ when $\Gamma=107$.
Note that  ${\tilde\mu}_0$ and $\Gamma$ are not independent parameters.
As shown in Fig. 2 our effective mass model with the paramagnetic 
effect explains the angle dependence of $H_{c2}(\theta)$ once we fix
one adjustable parameter.
It is to be noted as shown in inset of Fig. 2 the ${\tilde\mu}(\theta)$ value 
is completely determined by the effective mass form with $\Gamma=107$.

As for the assigned $\Gamma=107$ 
we point out that  the diamagnetic orbital current is determined by the perpendicular 
component of the average Fermi velocity to the field direction.
Thus $\Gamma$ is the
anisotropy ratio of the Fermi velocities, namely
$\Gamma=\sqrt{\langle v^2_{F\parallel c}\rangle/\langle v^2_{F\parallel ab}\rangle}$.
 This quantity is determined directly
by dHvA experiment; $\Gamma_{\alpha}=117$, $\Gamma_{\beta}=57$
and $\Gamma_{\gamma}=174$ for three bands $\alpha$, $\beta$ 
and $\gamma$ respectively~\cite{maeno0}.  Note that a simple geometric average
$\Gamma_{\rm eff}={1\over 3}(\Gamma_{\alpha}+\Gamma_{\beta}
+\Gamma_{\gamma})=116$ is well compared with our assignment $\Gamma=107$.
In this sense there is virtually no adjustable parameter in our analysis.
In passing we note that the observed ratio $\Gamma^{\rm obs}=
{H_{{c2}{\parallel ab}}/ H_{{c2}{\parallel c}}}=20$ is strongly reduced 
from $\Gamma_{\rm eff}$, apparently suggesting some reduction mechanism.
We clarified it here.

\begin{figure}[!h]
\includegraphics[width=6cm]{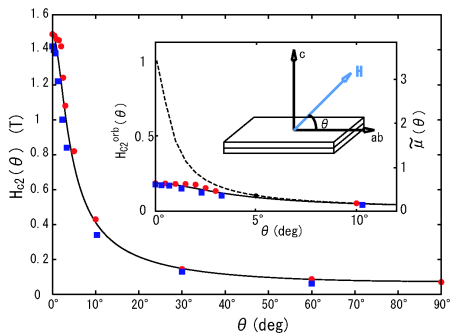}
  \caption{
(color online)
Calculated angle dependence of $H_{c2}(\theta)$ (solid line).
Circles~\cite{deguchi0} (squares~\cite{deguchi2}) are experimental data.
Enlarged figure is shown in inset for small angles. The 
dotted line is the original orbital limit $H^{\rm orb}_{c2}(\theta)$ of
the effective mass form with $\Gamma=107$.  
The dotted line also shows $\tilde{\mu}(\theta)$
with $\tilde{\mu}_0=3.41$ (right hand scale). $\theta$ is the
angle from the $ab$ plane.
}
 \label{fig:nchi}
\end{figure}

Let us now come to our main discussions on the analyses of 
the specific heat at a low $T$. In Fig. 3 we display $\gamma(H)$
for several values of ${\tilde\mu}$ together with 
the experimental data in inset (a) for various $\theta$ values.
They show strikingly similar behaviors as a whole.
The larger angle data exhibit a strong upward curvature, 
corresponding to the conventional $\gamma(H)\sim \sqrt H$
which is characteristic to the line node gap structure.
Those are reproduced in our ${\tilde\mu}$=0.02, or 0.41 curves.
As $\theta$ becomes smaller,
this changes into almost linear or concaved curves near $H_{c2}$.
This behavior is captured by the theoretical calculations for larger ${\tilde\mu}$'s.
Thus the overall ``metamorphosis'' of $\gamma(H)$ from
the conventional $\sqrt H$ to a strong convex curve is reproduced by increasing 
${\tilde\mu}$.
As shown in inset (b) of Fig. 3, the data are fitted well
by our calculations near $H_{c2}$ where 
we have used the ${\tilde\mu}(\theta)$ values determined 
above (see the inset of Fig. 2 with ${\tilde\mu}_0=3.41$).
We have computed  the six cases shown in Fig.3 for ${\tilde\mu}$ values 
and  obtained $\gamma(H)$ for other ${\tilde\mu}$'s by interpolation.

\begin{figure}[!h]
\includegraphics[width=7cm]{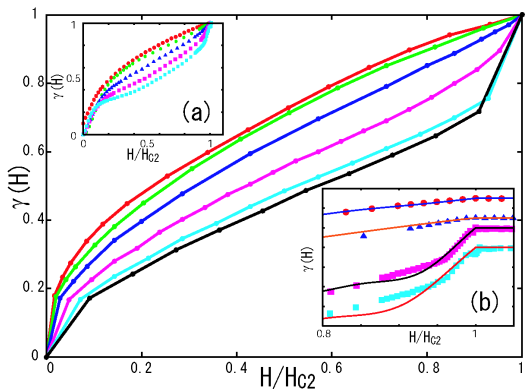}
  \caption{
(color online)
Zero-energy DOS $\gamma(H)$ at $T=0.1T_{\rm c}$  
for $\tilde{\mu}=0.02$, 0.41, 0.86, 1.71, 2.57 and 3.41
from top to bottom. 
Inset (a) shows the experimental data~\cite{deguchi0} for $\theta$=0$^{\circ}$, 
2.5$^{\circ}$, 3.0$^{\circ}$, 5.0$^{\circ}$ and 90$^{\circ}$ from bottom to top.
Inset (b) is the fitting of the data $\theta$=0$^{\circ}$
 by $\tilde \mu$=3.41, 0.5$^{\circ}$ ($\tilde \mu$=2.36), 
5$^{\circ}$ ($\tilde \mu$=0.33) and 90$^{\circ}$ ($\tilde \mu$=0.03) 
from bottom to top, which are shifted upwards.
}
\end{figure}

\begin{figure}[!h]
\includegraphics[width=\linewidth]{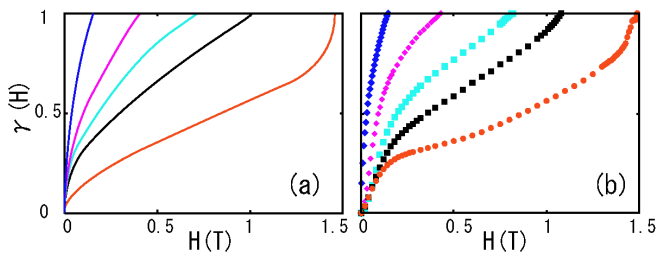}
  \caption{
(color online)
(a) $\gamma(H)$ for $\tilde{\mu}=3.41$, 0.60, 0.36, 0.18 and 0.06
from bottom to top.
(b) Corresponding data~\cite{deguchi0} for $\theta$=0$^{\circ}$, 3$^{\circ}$, 
5$^{\circ}$, 10$^{\circ}$ and 30$^{\circ}$.
}
\end{figure}

In Fig. 4 we display the theoretical $\gamma(H)$ behaviors (a)
and  the corresponding specific heat data~\cite{deguchi0} (b),
where we read off ${\tilde\mu}(\theta)$ from the inset of 
Fig.2.
Our theoretical curves explain
these data in a consistent manner. 
In particular, it is noteworthy;
(1) At $\theta$=0$^{\circ}$ where ${\tilde\mu}(0)={\tilde\mu}_0=3.41$
is largest, $\gamma(H)$ shows a $\sqrt H$-like sharp rise in smaller $H$ 
region because of the presence of line nodes. But it is limited only to lower fields.
(2) In the intermediate wide field region (0.5T$<H<$1T),
$\gamma(H)$ exhibits an almost linear change in $H$.
This extended linear change is shown to be consistent thermodynamically 
with magnetization $M(T,H)$ behavior as seen shortly.
(3) In the high field region ($H>$1T) towards $H_{c2}=1.5T$,
$\gamma(H)$ displays a sharp rise with a strong concave curvature.
As $H$ increases, the Pauli effect proportional linearly to $H$ becomes growingly
effective, modifying $\gamma(H)$ from usual $\sqrt H$
to a concave $H^{\alpha}$-like curve with $\alpha>$  1.

The data for $\theta$=3$^{\circ}$ where ${\tilde\mu}(\theta=3^{\circ})=0.60$
show a similar behavior to that at $\theta$=0$^{\circ}$, but the 
features associated with the Pauli effect, 
namely, the existence of the inflection point from
convex to concave curves  and sharp rise towards $H_{c2}$ are weaken.
The $\gamma(H)$ data for higher angles ($\theta>3^{\circ}$)
exhibit an intermediate behavior between those at $\theta=0^{\circ}$ and the
ordinary $\sqrt H$ curve, continuously changing its shape with 
$\theta$.  It is remarkable 
that the strong concaved curves of the experimental
data for small angles, which were unexplained before, are
reproduced by the Pauli paramagnetic effect.
Physically, this effect makes the conventional Abrikosov vortex state
unstable, ultimately leading to the normal state
via a first order transition or the FFLO state.
The sharp rise in $\gamma(H)$ near $H_{c2}$ is a precursor to it.

In Fig. 5 we show the calculated results of magnetization $M(H)$
for several $T$'s (a) together 
with the experimental data\cite{tenya} (b)
to qualitatively understand the paramagnetic effects on $M(T, H)$. 
We do not attempt to reproduce the data quantitatively because the data 
are in a qualitative nature due to hysteresis effects.
It is seen from Fig. 5(a) that the magnetization with
a convex curvature at lower field changes into that with a concave one towards 
$H_{c2}$. There is an inflection point field $H_{K}$ in between.
The relative position of $H_{K}$ to each $H_{c2}$ 
decreases with $T$ (also see insets). In higher $T$'s $H_{K}$ becomes invisible 
because of thermal effect. These two features are observed experimentally as
seen from Fig. 5(b).  The inflection point field $H_{K}$
 roughly coincides with that in $\gamma(H)$
as seen from Fig. 4, implying that these are thermodynamically related
to each other.

As is seen from Fig.5 upon lowering $T$ the slope of $M(H)$ at 
$H_{c2}$ becomes steeper, meaning that $\kappa_2$ decreases,
instead of increases as in usual superconductors\cite{saint}.
This is another obvious supporting evidence that the paramagnetic
effect is important in Sr$_2$RuO$_4$.

\begin{figure}[!h]
\includegraphics[width=\linewidth]{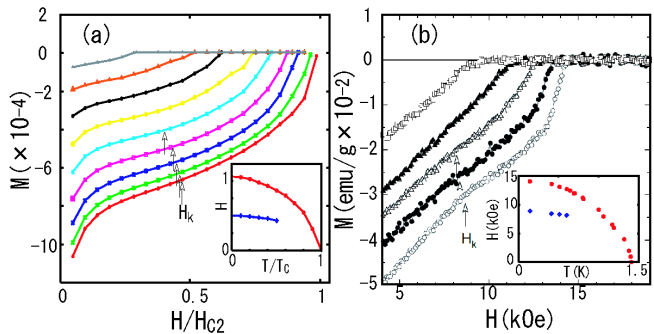}
  \caption{
(color online)
(a) Calculated magnetization curves for various $T/T_c=
0.1, 0.2, 0.3\cdots, 0.9$ from bottom to top for $\tilde{\mu}=1.71$. 
Inset shows $H_{c2}$ and the inflection point $H_K$.
(b) Corresponding data~\cite{tenya} for $T/T_c$=0.1, 0.28, 0.40 and 
0.56 from bottom to top for $H\parallel ab$. Inset shows
$H_{c2}$ and ``kink" field $H_K$ in their terminology~\cite{tenya}.
Magnetization of the normal paramagnetic moment is substracted.
}
\end{figure}

It is easy to derive a thermodynamic Maxwell relation 
${d\over dH}{C\over T}={\partial^2\over \partial T^2}M(T,H)$
from which we can see at low $T$,
${\partial \gamma(H)\over \partial H}=\beta(H)$
with $M(T,H)=M_0(H)+{1\over 2}\beta(H)T^2$.
We estimate $\beta(H)$ from the experimental data~\cite{tenya}
in Fig. 5, finding that $\beta(H)\sim$ const for 0.5T$<$$H$$<$1T and 
$\beta(H) \propto H^3$ for 1T$<$$H$$<$1.35T. This implies that
$\gamma(H)\propto H (H^4)$ for 0.5T$<$$H$$<$1T (1T$<$$H$$<$1.35T).
These behaviors in $\gamma(H)$ are indeed seen
for the $\theta=0^{\circ}$ data shown in Fig. 4.
These analyses, which are free from any microscopic model, mean
that the mysterious behavior of $\gamma(H)$ is supported 
to be true thermodynamically
and comes from the intrinsic nature deeply rooted to the
superconductivity in Sr$_2$RuO$_4$.

There are several known difficulties associated with
the most popular two component chiral $p$-wave pairing;
${\hat z}(p_x+ip_y)$\cite{sigrist} or ${\hat z}(p_x+ip_y)\cos p_z$\cite{hasegawa}:
Experimentally these triplet states are unable to explain the paramagnetic
effects mentioned above because the $d$-vector is not locked  in the basal plane.
Theoretically these states give a large in-plain $H_{c2}$ anisotropy\cite{kauer}
which is not observed. The present singlet scenario is free from it. 

Let us go on considering the high field 
phase for $H$$\parallel$$ab$ observed as the double transition\cite{deguchi2}.
It appears in a narrow $H$-$T$ region along $H_{c2\parallel ab}$,
starting at $T_0=0.8K$, or $T_0=0.53T_c$
at which three transition lines meet, giving rise to
a tricritical point in $H$ vs.. $T$ plane.
$T_0$ is remarkably similar to the so-called Lifshitz point 
$T_L=0.56T_c$ in the FFLO phase diagram for a 
Pauli limited superconductor where the orbital depairing is quenched
completely. This number $T_L=0.56T_c$ is universal, valid for a 
variety of situations, including 3D Fermi sphere $s$-wave\cite{takada},
2D $s$-wave\cite{shimahara} and $d$-wave\cite{sauls2}, 
and 1D $s$-wave\cite{nakanishi} models.
Our identified large paramagnetic parameter $\tilde\mu=3.41$
means that our system is in almost Pauli limiting
where the orbital effect is almost perfectly quenched because the two-dimensionality
in Sr$_2$RuO$_4$ is so extreme.
In fact note that the identified anisotropy $\Gamma$=107 implies
$H_{c2\parallel ab}^{\rm orb}\sim7.5T$ which is reduced to  
$H_{c2\parallel ab}=1.5T$ by the Pauli effect. 
Thus we propose here to identify this high field phase as FFLO.

The extreme two-dimensionality is obvious:
If $H$ is tilted away from the $ab$ plane only by $\theta>0.3^{\circ}$,
the double transition vanishes\cite{deguchi2}. According to Nakai, {\it et al.}\cite{nakai}
the FFLO region at low $T$ occupies $\sim 0.8\%$
below $H_{c2}$, which is comparable with the width $\sim$200G
of the high field phase below $H_{c2\parallel ab}=1.5T$, a region
200G/1.5T$\sim$1.3\%\cite{deguchi2}.
Guided by the known phase diagram\cite{saint},
we predict that as the field orientation $\theta$ increases,
$\tilde \mu$ decreasing, this high field phase survives only for 
$0<\theta<0.3^{\circ}$ and quickly diminishes for $\theta>0.3^{\circ}$. 
At around $\theta\sim 1.0^{\circ}$
there appears a first order transition along $H_{c2}$ line instead of FFLO.
Then for $\theta>2.0^{\circ}$ it also disappears above which
the paramagnetic effect becomes ineffective and 
Sr$_2$RuO$_4$ is described by a conventional singlet
superconductor with line nodes. These predictions 
based on our analyses are all testable experimentally 
although the details should be further sharpened theoretically.

In conclusion, we have analyzed both specific heat at lower $T$
and magnetization $M(T,H)$ by self-consistently solving microscopic 
quasi-classical Eilenberger equation 
for the gap function with line nodes. It is seen that the Pauli paramagnetic depairing
effect is essential in understanding the data in Sr$_2$RuO$_2$.
This is possible only for either singlet pairing,
or triplet pairing with the d vector locked in the basal plane.

We are grateful for useful discussions and communications to K. Deguchi, K. Tenya,
K. Ishida, Y. Maeno, H. Adachi, N. Nakai, P. Miranovic and Y. Matsuda. 
M. Tsutsumi helps us for preparing figures.

\end{document}